\documentstyle[aps,pre]{revtex}

\newcommand \bea {\begin{eqnarray}}
\newcommand \eea {\end{eqnarray}}
\newcommand \be {\begin{equation}}
\newcommand \ee {\end{equation}}

\newcommand \bi {\bibitem}

\begin{document}

\title{A glass transition scenario based on
 heterogeneities and entropy barriers}

\author{Andrea Crisanti$^{1}$ and 
 Felix Ritort$^{2}$} \address{$^{1}$Dipartimento di Fisica, Universit\`a di Roma ``La Sapienza'',
         P.le Aldo Moro 2, I-00185 Roma, Italy \\
         Istituto Nazionale Fisica della Materia, Unit\`a di Roma}
 
\address{$^{2}$Departament de F\'{\i}sica Fonamental,
 Facultat de F\'{\i}sica, Universitat de Barcelona\\ Diagonal 647,
 08028 Barcelona (Spain)\\ E-Mail: crisanti@phys.uniroma1.it,ritort@ffn.ub.es}


\maketitle
\begin{abstract}
We propose a scenario for the glass transition based on the
cooperative nature of nucleation processes and entropic
effects. The main point is the relation between the off-equilibrium energy 
dissipation and nucleation processes in off-equilibrium supercooled liquids
which leads to a natural definition of the complexity. From the absence of 
coarsening growth we can derive an entropy based fluctuation formula which 
relates the free energy dissipation rate in the glass with the nucleation rate 
of the largest cooperative regions.  
As by-product we obtain a new phenomenological relation between the largest
relaxation time in the supercooled liquid phase and an effective
temperature. This differs from the Adam-Gibbs relation in that predicts
no divergence of the primary relaxation time at the Kauzmann temperature and 
the existence of a crossover from fragile to strong behavior.
\end{abstract}

\section{Introduction}
Standard approaches to the glass transition have been
largely based on hydrodynamic or thermodynamic entropic
theories like that proposed by Adam, Gibbs and Di Marzio (hereafter
referred to as AGM) nearly 50 years ago \cite{AGM}. Surprisingly, still nowadays
the ideal AGM theory remains not accepted nor disproved,
and an unifying description of the glass transition problem is still not
available. A striking outcome of the AGM theory is the prediction of 
a second-order phase transition driven by the collapse of the
configurational entropy (also called complexity) at the Kautzman temperature
$T_K$. 

Strictly speaking an unambiguous definition of the complexity
is only possible in mean-field theories where phase
space splits into ergodic components of infinite lifetime. Recent
approaches to the glass transition problem from the perspective of
disordered systems have, to a large extent, validated this mean-field
scenario \cite{KTW}. Nevertheless, a complete understanding of the glass
transition must go beyond mean field by including nucleation processes
into any valuable theory. This poses the question about how the
mean-field picture for the glass transition is modified in the
presence of real-space effects. 

Here we propose a scenario for the glass transition
where spatial effects due to heterogeneities play the central role.

\section{Phenomenological description of heterogeneities in glasses}
One of the most intriguing features in glasses is the existence of
heterogeneous structures \cite{HETERO}. Experimentally these manifest
as some set of atoms which have a dynamics manifestly slow when
compared to the rest. Although experiments or numerical simulations in
this field are very recent, heterogeneities are a direct manifestation
of the cooperative nature of nucleation processes \cite{IS}. 
The heterogeneities observed in the experiments are
transient frozen structures which nucleate on time scales of the order or 
larger than the observational time. 
When a droplet of liquid nucleates it changes to a new locally disordered 
structure. The local structure of the glass is always that of a liquid and 
no coarsening of a given pattern is observed.

Our starting point is tha asspumption that nucleation processes take place
everywhere inside the glass when some correlated structures of typical
size $s$ are built by an anchorage or cooperative mechanism. By
anchorage mechanism we mean that $s$ atoms must coherently move to
occupy positions which enable that region to release some stress
energy. Each of these moves constitutes an elementary activated
process, for instance, the exchange between two neighboring
particles. In our scenario the time to anchor a region of
size $s$ is,
\be
\frac{\tau(s)}{\tau_0}\propto \left(\frac{\tau^*}{\tau_0}\right)^s=
\exp\left(\frac{Bs}{T}\right)
\label{eqxi}
\ee
where $\tau^*=\tau_0\exp(B/T)$ is the activated time to anchor one
atom, $\tau_0$ being a microscopic time and $B$ the corresponding
energy barrier. 

Experimentally it is
known that time correlations in the glass state are stretched but
still decay faster than any power law. Because of (\ref{eqxi}) this
means that the distribution $n_s(t)$, the number of cooperative regions
(domains)
of size $s$ which have already nucleated at time $t$, 
must fall down abruptly beyond a maximum size $s^*(t)$ 
for extremely long relaxations to occur with a negligible
probability. Consequently, $n_s(t)$ must have a well defined cuttof at
$s^*$ such that $n_s(t)$ for $s>s^*$ is nearly zero
\cite{WX}.  It can be shown that the existence of this cutoff is
tightly related to the cooperative character of the dynamics itself
eq.(\ref{eqxi}) and can be illustrated within a simple domain
aggregation model \cite{CR01}. 

Consequently, relaxation to equilibrium of, say, a supercooled liquid
is driven by the growth of the largest domain size or cutoff value $s^*(t)$ 
since the release of energy to the thermal bath occurs
only when the largest domains of size $s^*$ nucleate for the first time.
Nucleation processes of regions of size smaller than $s^*(t)$
always occur but do not yield a net thermal current flow to the thermal
bath. Metastable equilibrium is reached when the cutoff size $s^*$
saturates and the net energy flow between the glass and the bath vanishes.

Without knowing any further details about the hydrodynamic equations
which describe the relaxation we can set up a phenomenological equation for the
energy dissipation inside the glass. This may be described by any
thermodynamic potential variable, for instance the free energy
$F$. The release of energy per unit time inside the glass is given by
typical free energy variation of those domains which relax (i.e those
with size $s^*(t)$) multiplied by its multiplicity $N_{s^*}(t)=V
n_{s^*}(t)=(V/s^*)\hat{n}(1)$ \cite{CR01}, divided by their typical nucleation
time $t^*=\exp(B\beta s^*)$,
\be
\frac{1}{V}\frac{\partial F}{\partial t}\sim \frac{\Delta
F^* N_{s^*}(t)}{Vt^*}\sim \frac{\Delta F^*}{s^* t^*}~~~~.
\label{dissip1}
\ee

The crucial point which distinguishes nucleations processes in
glasses from other physical systems is the fact that nucleation occurs
between liquid drops. This means that contrarily to standard nucleation theory, 
the typical release of energy
$\Delta F^*$ by nucleating a liquid droplet of size $s^*$ does not
scale with its surface (like in coarsening systems) nor its volume
(like in standard liquid-solid nucleations) but is finite and
independent of $s^*$ \cite{AGM} yielding,
\be
\frac{1}{V}\frac{\partial F}{\partial t}\sim-\frac{\Delta_F}{s^* t^*}~~~~.
\label{dissip1_b}
\ee
where $\Delta_F$ is the average free energy fluctuation.
The size $s^*$ grows with time as $s^*(t)\sim T\log(t)$ 
and stops when $n_s(t=t_{\rm eq})$ reaches the
stationary distribution. In principle the value of the equilibration
time $t_{\rm eq}$ depends on the particular model under
consideration. From a microscopic theory this information is only
contained in the hydrodynamic equations. We will see below how we can
circumvent the hydrodynamic description by relating the relaxation
time of the supercooled liquid to the complexity density.

\section{The fluctuation formula.} 
One of the main problems of the
standard AGM theory is how to define the complexity $S_c$. In the
original AGM theory, $S_c$ was defined as the number of conformational
states of the liquid compatible with a given energy. The problem in
this definition is the following: if the complexity is a static
quantity then one is naturally led to the conclusion that it vanishes
in the supercooled liquid line if measured over time scales much larger
that the largest decorrelation time.  In turn, according to the
Adam-Gibbs relation, this implies a divergence of the relaxation
time. This contradiction can be avoided defining a dynamical
complexity $S_c(t^*,F)$ by counting all conformations (or basins) of
the liquid with free energy $F$ which nucleate in a time of the order
of the primary relaxation time. Recently, this definition has been
applied to some mean-field spin glass models where nucleation
processes are neglected \cite{THEO,KB}.  Let us assume that the
liquid is in off-equilibrium after a quench  to temperature 
$T_f$. Here we define $S_c(t^*,F)$ as the logarithm of the total
number of conformations with free energy $F$ which can release energy
after nucleating for their first time at time $t^*$. In other words,
$S_c(t^*,F)$ counts the number of still {\it not visited}
conformations at time $t^*$. In this scenario regions which have
nucleated once have already released their energy while regions which
have not nucleated still contain some strain energy. Nucleations which
do not lead to a release of strain energy do not yield new
conformational states.

Each of these conformations at
time $t^*$ may contain different possible configurations ${\cal C}$
which do not contribute to the complexity but contribute to the free
energy of the conformation ${\cal B}$, $F_{\cal
B}(t^*,T_f)=-T_f\log(\sum_{{\cal C}\in {\cal B}} \exp(-E({\cal
C})/T_f)$ allowing also to define time dependent expectation values of
any observable $A_{\cal B}(t^*,T_f)$ as follows: $A_{\cal
B}(t^*,T_f)=\sum_{{\cal C}\in{\cal B}}A({\cal
C})\exp\Bigl(-\frac{(E({\cal C})-F_{\cal B}(t^*,T_f))}{T_f}\Bigr)$.

The liquid character of the glass phase implies that basins with
identical free energy $F_{B}(t^*,T_f)$ have 
the same physical properties.  Note that the free energy and the
expectation value of any observable $A$ evaluated at a given
conformation ${\cal B}$ depend on both $T_f$ and the quenching time
$t^* $through the configurations ${\cal C}$ contained in ${\cal B}$ 
In particular, note that for large times basins with
very high free energy only contain configurations which have nucleated
several times, hence they do not contribute to the
complexity. Contrarily, conformations with very low free energy
contain configurations which have still not nucleated so they
contribute to the complexity. This leads to a dynamical complexity
having a time-dependent cutoff value $F^*$ such that the number of
conformations or basins with $F_{\cal B}(t^*,T_f)>F^*$ vanishes.  The
complexity is a monotonous increasing function of the free energy
since the number of possible conformations which decrease the energy
is larger when nucleating regions are smaller. In the asymptotic large
$t^*$ limit it takes the form,
\be
S_c(t^*,F<F^*)=\hat{S_c}(F,T_f)~;~S_c(t^*,F>F^*)\to -\infty
\label{Sc}
\ee

Since at time $t^*$ after the quenching nucleations inside the glass
occur between cooperative regions with a local liquid structure and no
coarsening process drives the nucleation, nucleations must be
entropically driven. In other words, the probability to jump from any conformation
with free energy $F$ onto a conformation 
with free energy $F'$ is always proportional to the number
of configurations with final free energy $F'$,

\be {\cal W}_{F\to F'}\propto \frac{\Omega(F')}{\Omega(F)}\propto
\exp(\hat{S_c}(F',T_f)-\hat{S_c}(F,T_f))~~~.
\label{W}
\ee

Substituting (\ref{Sc}) in (\ref{W}) and denoting $\delta F=F'-F$ 
we obtain for the transition probability,

\be
{\cal W}_{\delta F}\propto\exp\Bigl[\beta_{\rm eff}(F^*)\delta
F\Bigr]\theta(-\delta F)
\label{FF}
\ee

where we have defined the effective temperature $\beta_{\rm
eff}(F^*)=\bigl(\frac{\partial \hat{S_c}(F,T_f)}{\partial F}\bigr)_{F=F^*}$ 
as the density of complexity with free energy $F^*$.  
This is the fluctuation formula which
establishes the probability of free energy fluctuations. Fluctuations to
conformations or basins which increase the free energy are forbidden,
simply because they have already nucleated in the past. While
fluctuations to very low free energy conformations are entropically
suppressed due to the monotonous increasing property of $\hat{S_c}(F,T_f)$.

Using the above expression for the free energy fluctuations it follows that
in the off-equilibrium state the average rate variation of the free
energy at time $t^*$ is then given by,
\be \frac{1}{V}\frac{\partial F}{\partial t}\sim
-\frac{\int_{-\infty}^{\infty}x{\cal
W}_xdx}{t^*\int_{-\infty}^{\infty}{\cal W}_xdx}= -\frac{1}{\beta_{\rm
eff}(F^*) t^*}.
\label{dissip1_c}
\ee

Consistency between this equation with the hydrodynamic equation
(\ref{dissip1_b}) requires that 
$s^*=\Delta_F\beta_{\rm eff}(F^*)$, i.e. 
the size of the largest nucleating regions $s^*$ is
directly proportional to the complexity density (also called effective
temperature) $\beta_{\rm eff}(F^*)$ evaluated at the time-dependent
free energy $F^*$.  

From the scenario drawn above off-equilibrium relaxation goes as follows: as
time goes on, $F^*$ decreases up to time $t_{\rm eq}$ when
it converges to the equilibrium value $F_{\rm eq}=F(t_{\rm
eq},T_f)$. At this time, $s^*$ has saturated to a value 
$s^*(t_{\rm eq})$ which determines the phenomenological relation,
\be
\tau_{\alpha}=\tau_0\exp(Bs^*(t_{\rm
eq})/T_f)=\tau_0\exp(B\Delta_F\beta_{\rm eff}(F_{\rm eq})/T_f).
\label{CR}
\ee

In the AGM scenario \cite{AGM}
the number of different conformations at time $t^*$
corresponds to all possible combinations $\Omega_{t^*}$ obtained from
the two possibilities (nucleated or not nucleated) for the different
independent largest nucleating regions $n_{s^*}(t^*)$. The number of
conformations is given by $\Omega_{t^*}=2^{\frac{V}{s^*}}$ yielding
for the complexity, $S_c(t^*)=\log(\Omega_{t^*})=V(\log(2)/s^*)$
directly relating the size of the cooperative region to the
complexity. Using eq.(\ref{eqxi}) and taking $t^*=t_{\rm eq}$ this
yields the famous Adams-Gibbs relation
$\tau_{\alpha}=\exp(B\log(2)/T_fs_c)$. By analyzing in details the AGM
scenario we see that it contains the strong assumption that 
in the off-equilibrium regime the glass explores conformations differing 
by the nucleation of the largest regions as if they were in equilibrium at 
the quenching temperature $T_f$, assumption which obviously cannot hold if 
the system is in off-equilibrium and free energy fluctuations are biased 
towards lower free energy conformations. Note that according to AGM, the size 
of the cooperative region diverges at the Kauzmann temperature
$T_K$ while in the scenario presented here it saturates to a finite value.

\section{Fluctuation-Dissipation relation}
One more consequence of the fluctuation formula (\ref{FF}) concerns the
fluctuation-dissipation ratio (FDR) and its one-step
character \cite{CK,SIM}. After quenching to $T_f$ a possible way to quantify
violation of FDR is to measure the average value of any observable $A$
after a perturbation field $h_A$ conjugated to the observable $A$ is
applied to the system. Due to the liquid order of the cooperative
regions, if the perturbation does not bias one class of conformation
over others, the entropically driven assumption implies that states with
free energy $F$ are sampled with a probability according to their
number. In the presence of a field $h_A$ the complexity must be a
function of three variables, $\hat{S_c}(F,T_f,A)$. We can simply obtain the
average change in the expectation value $<A(t)>$ after switching on the
perturbation field $h_A$ at time $t_w$. In the linear response regime
\cite{FV} the Onsager regression principle implies for the transition
probabilities 
\begin{eqnarray}
{\cal W}_{\delta F,\delta A=A-A_0}&=&
    {\cal W}_{\delta F}\exp[(\hat{S_c}(F,T_f,A)-\hat{S_c}(F,T_f,A_0)]
\nonumber \\
                                  &=&{\cal W}_{\delta
F}\exp\left(\frac{\partial \hat{S_c}(F,A)}{\partial A}\right)_{A_0}.
\end{eqnarray}
Using the relation
$(\frac{\partial \hat{S_c}(F,A)}{\partial A})_{A_0}=-\beta_{\rm eff}(F)h_A$
and expanding for $h_A$ small we finally get the famous violation FDT
expression, 
\begin{equation}
\frac{\partial \langle A\rangle (t)}{\partial h_A(t_w)}=\beta_{\rm
eff}(F^*)(\langle A(t)(A(t)-A(t_w))\rangle _{h_A=0}.
\end{equation}
The description of the
violation of FDT in terms of a single time scale $t^*$ is therefore consequence of
the asymmetric shape (i.e. ${\cal W}_{\delta F>0}=0$) of (\ref{FF}).

\section{Main implications of the present scenario.} 
It can be proven \cite{MARC} that $\beta_{\rm eff}(F)\le 1/T_K$ where
$\hat{S_c}(T_K)=0$ at the Kauzmann temperature $T_K$. This implies an
asymptotic crossover for all fragile liquids to strong behavior. A
strong glass is a fragile one which has exhausted all its complexity
and the effective barrier has saturated to its maximum value
$\beta_{\rm eff}=1/T_K$. Instead fragile glasses have high excess
complexity and still big variation of the complexity density
$\beta_{\rm eff}$ along the supercooled line. The dependence of the
activation barrier in (\ref{CR}) through the non-universal quantity
$\beta_{\rm eff}$ explains why it is so difficult to find a unique
empirical law that properly describes the viscosity anomaly of all
glasses. The prediction that $s^*$ saturates to a finite value
$s^*=\Delta_F/T_K$ at $T_K$ is not easy to check experimentally due to
the difficulty to cover one order of magnitude in the effective
barrier. Despite that experimental results suggest a crossover from
fragile to strong crossover scenario \cite{CAVAGNA,CROSSOVER} the present
scenario can be also checked doing numerical aging
experiments. According to relation (\ref{CR}) the $t_w$ dependent
effective temperature $\beta_{\rm eff}(t_w)$ (measured through
FDT-violations or the formula \ref{FF}) should be given by $\beta_{\rm
eff}(t_w)\to(T_f/B\Delta_F)\log(t_w/\tau_0)$ if
$\tau_{\alpha}(t_w)\simeq t_w$ and both $\tau_0$ and $B\Delta_F$ are
nearly $t_w$ and $T_f$ independent. This relation should hold for all
$T_f$ and $t_w$ predicting a value for the activation barrier and the
cooperative size \cite{ROM}. The experimental confirmation of a
saturation of the heterogeneity sizes for not too fragile glasses and
the crossover from fragile to strong behavior is probably not out of
reach and would be a check of the validity of the present theory.

{\bf Acknowledgments}. We wish to thank suggestions by C. Cabrillo, G. Parisi,
F. Sciortino and G. Tarjus. F.R. is
supported by the MEC in Spain, project
PB97-0971.
\hspace{-2cm}

\end{document}